\documentclass[aps,prc,twocolumn,groupedaddress]{revtex4}

\usepackage[utf8]{inputenc}  
\usepackage{color}
\usepackage{graphicx}
\usepackage{mathrsfs}
\usepackage{bm}
\usepackage{amsmath,amssymb,amsfonts}
\usepackage{amssymb}
\usepackage{ulem}
\usepackage{dsfont}
\usepackage{adjustbox}

\makeatletter

\providecommand{\tabularnewline}{\\}

\makeatother

\begin{document}

\title{Neutron scattering off spherical nuclei with global nonlocal dispersive optical model}

\author{B.~Morillon}\email{benjamin.morillon@cea.fr}
\affiliation{CEA,DAM,DIF F-91297 Arpajon, France}
\affiliation{Universit\'e Paris-Saclay, CEA, LMCE, 91680 Bruyères-le-Ch\^atel, France}

\author{G.~Blanchon}
\affiliation{CEA,DAM,DIF F-91297 Arpajon, France}
\affiliation{Universit\'e Paris-Saclay, CEA, LMCE, 91680 Bruyères-le-Ch\^atel, France}

\author{P.~Romain}
\affiliation{CEA,DAM,DIF F-91297 Arpajon, France}
\affiliation{Universit\'e Paris-Saclay, CEA, LMCE, 91680 Bruyères-le-Ch\^atel, France}

\author{H.~F.~Arellano}
\affiliation{Department of Physics-FCFM, University of Chile, Av. Blanco Encalada 2008,
  RM 8370449 Santiago, Chile}
\affiliation{CEA,DAM,DIF F-91297 Arpajon, France}

\date{Received: date / Revised version: date}

\begin{abstract}
  We present a global nonlocal and dispersive optical model potential for
  neutron scattering off spherical nuclei with incident energies up to 250 MeV.
  This optical model is an extension of the non-dispersive Perey-Buck potential.
  The imaginary components are chosen energy-dependent and the dispersive
  constraints are taken into account. The surface imaginary part is nonlocal, 
  whereas the volume imaginary part above 10~MeV is local, allowing to
  reproduce total cross sections and scattering data for high energies.
  We obtain a good description of scattering observables for target-nuclei
  ranging from $A=16$, up to $209$. The inclusion of nonlocal spin-orbit term enables
  a better description of the analyzing power data relative to the local dispersive
  model. 
\end{abstract}

\maketitle

\section{Introduction}
\label{sec::intro}

More than fifteen years ago, two of us introduced a global local nucleon Optical
Model Potential (OMP) \cite{morillon_04,morillon_06,morillon_07} including dispersion relations
\cite{mahaux_91} and the local energy approximation of Perey-Buck \cite{perey_62}.
This potential provides a good description of integral and differential elastic cross sections for
neutron and proton scattering off spherical target-nuclei for incident energy up
to 250~MeV. Bound state energies were used as constraints for the calibration of the
potential for neutrons \cite{morillon_06} and protons \cite{morillon_07}. The calibration
of the parameters was made using the same data set of Koning-Delaroche (KD) in Ref.~\cite{koning_03},
i.e. mostly stable and spherical target-nuclei. Based on the same method,
Charity \textit{et al.} have studied the evolution of the OMP for target-nuclei with
large proton-neutron asymmetry \cite{charity_06,charity_07,mueller_11}. 

Since then, various attempts have been pursued to handle nonlocality explicitly.
Tian \textit{et al.} have extended Perey-Buck (PB) potential \cite{perey_62} to include
both neutron and proton projectiles \cite{tian_15}. More recently, Mahzoon~\textit{et~al.}
have proposed a dispersive nonlocal potential \cite{mahzoon_14}. This potential has allowed
for the first time to reproduce the target-density and the particle number of the target-nucleus
in addition to the usual scattering observables. First applied to $^{40}$Ca for both neutron and
proton projectiles, the approach has been extended to include $^{48}$Ca \cite{mahzoon_17}
and $^{208}$Pb \cite{atkinson_20}. In parallel, Lovell~\textit{et al.} \cite{jaghoub_18} and
Jaghoub \textit{et al.} \cite{lovell_17} have studied the energy dependence of nonlocal
potentials. Those studies are all based on PB separable potential which includes a Gaussian
form factor for the nonlocality \cite{perey_62}.
In a recent review, Hebborn~\textit{et al.} summarize efforts toward improvements, within
both phenomenological and microscopic optical models approaches \cite{hebborn_23}. 

In this work, we present the first global dispersive and nonlocal OMP for neutron scattering
off spherical target-nuclei with incident energy up to 250~MeV. We adopt the PB form factor
as starting point. The imaginary contribution to the OMP is taken energy-dependent whereas
the real contribution is energy dependent only through the dispersion relation. The account
for nonlocality enables us to get rid of spurious energy dependence stemming from local
approximations. The calibration of the OMP parameters is done for twenty-two nuclei with
masses ranging from $A=16$, up to $209$. Both experimental elastic scattering observables and
bound state energies are considered in the calibration process. 
\\

We organize this work as follows. In Sec.~\ref{sec::perey}, we recall basics of Perey-Buck
potential model. Then we contrast the predictive power of the two most common parametrizations:
PB \cite{perey_62} and Tian, Pang and Ma (TPM) \cite{tian_15}. In Sec.~\ref{sec::nldisp}, we
present form factors for the real and imaginary components of the new nonlocal dispersive OMP.
The resulting parameters of this OMP are presented and discussed in Sec.~\ref{sec::opt}. Comparisons
are made in Sec.~\ref{sec::comp_local_nl} between calculated and experimental neutron cross sections,
analyzing powers and bound single-particle neutron states. Finally, in Sec.~\ref{sec::conc} we
present the main conclusions of this work. An appendix is added where we expose the method
used to solve the integro-differential Schr\"{o}dinger equation is exposed in
Appendix~\ref{sec::wf-calc}.

\section{Perey-Buck optical model}
\label{sec::perey}

In the early sixties, Perey and Buck proposed a nonlocal phenomenological potential to
describe neutron scattering off nuclei with incident energies below 28~MeV \cite{perey_62}.
The central part of PB potential reads
\begin{eqnarray}
  V(\mathbf{r},\mathbf{r}')=U\left(\frac{\left|\mathbf{r}+\mathbf{r}'\right|}{2}\right)
  H(\left|\mathbf{r}-\mathbf{r}'\right|),
  \label{eq::pb}
\end{eqnarray}
where $H$ represents the nonlocality form factor assumed Gaussian which is given as
\begin{eqnarray}
  H(\left|\mathbf{r}-\mathbf{r}'\right|)=\frac{1}{\pi^{\frac{3}{2}}\beta^{3}}
  \exp{\left[-\frac{(\mathbf{r}-\mathbf{r}')^{2}}{\beta^{2}}\right]}.
  \label{eq::pbnl}
\end{eqnarray}
Here $\beta$ is the range of the nonlocality. The radial form factor $U$ is given
the Woods-Saxon form
\begin{eqnarray}
  U(\tilde{r})=V_{V}f(\tilde{r},R,a)-i W_{S}4a\frac{df(\tilde{r},R,a)}{d\tilde{r}},
  \label{eq::rad-ff}
\end{eqnarray}
with $f$ a two-parameter Fermi distribution given by
\begin{eqnarray}
  f(\tilde{r},R,a)=\frac{1}{1+e^{\frac{{\textstyle \tilde{r}-R}}{{\textstyle a}}}}.
\end{eqnarray}
Here $\tilde{r}= \frac{1}{2}\left|\mathbf{r}+\mathbf{r}'\right|$, $a$ is the diffuseness parameter
and $R=r_{0}A^{1/3}$ is the target-nucleus radius, with $r_{0}$ the reduced radius and $A$ the
nucleus mass. Additionally the real volume strength $V_{V}$ and the imaginary surface
strength $W_{S}$ are energy independent. No imaginary volume component was introduced
in PB between 400~keV and 28~MeV. In addition, PB adopted a local prescription for the
spin-orbit potential.

When dealing with nonlocal potentials, the scattering equations turn to be 
integro-differential. One can express then scattering equations in partial waves
\begin{widetext}
\begin{eqnarray}
  \frac{\hbar^{2}}{2\mu}\left[\frac{d^{2}}{dr^{2}}
  -\frac{l(l+1)}{r^{2}}\right] u_{jl}(r) + E u_{jl}(r) 
  -V_{jl}^{\mathrm{L}}(r) u_{jl}(r)
  -\int_{0}^{\infty}\nu_{jl}^{\mathrm{NL}}(r,r^{\prime})u_{jl}(r^{\prime})dr^{\prime}=0,
  \label{eq::intdiff}
\end{eqnarray}
\end{widetext}
with $u_{jl}(r)$ the radial wave function and $\mu$ the projectile-target reduced
mass \cite{joachain_75}. For the sake of completeness, we include both local and nonlocal
contributions in Eq.~\eqref{eq::intdiff}. In the case of PB, the spin-orbit potential
is local, where
\begin{eqnarray}
V_{jl}^{\mathrm{L}}(r)=-(j(j+1)-l(l+1)-3/4) U_{so}(r),
\end{eqnarray}
which makes the local contribution $(j,l)$ dependent. The form factors in Eqs.~\eqref{eq::pb}
and \eqref{eq::pbnl} allow for analytical expressions for the potential multipoles,
\begin{eqnarray}
  \nu_{l}^{\mathrm{NL}}(r,r^{\prime})=
  \frac{4rr^{\prime}}{\sqrt{\pi}\beta^{3}}U\left(\frac{r+r^{\prime}}{2}\right)
  e^{{\textstyle -\frac{(r^{2}+r^{\prime2})}{\beta^{2}}}}i^{l}j_{l}\left(-i\frac{2rr^{\prime}}{\beta^{2}}\right).
  \nonumber\\
\end{eqnarray}
The method used to solve Eq.~\eqref{eq::intdiff} is described in Appendix~\ref{sec::wf-calc}.
\\

\begin{figure}
  \centering{\includegraphics[scale=0.55]{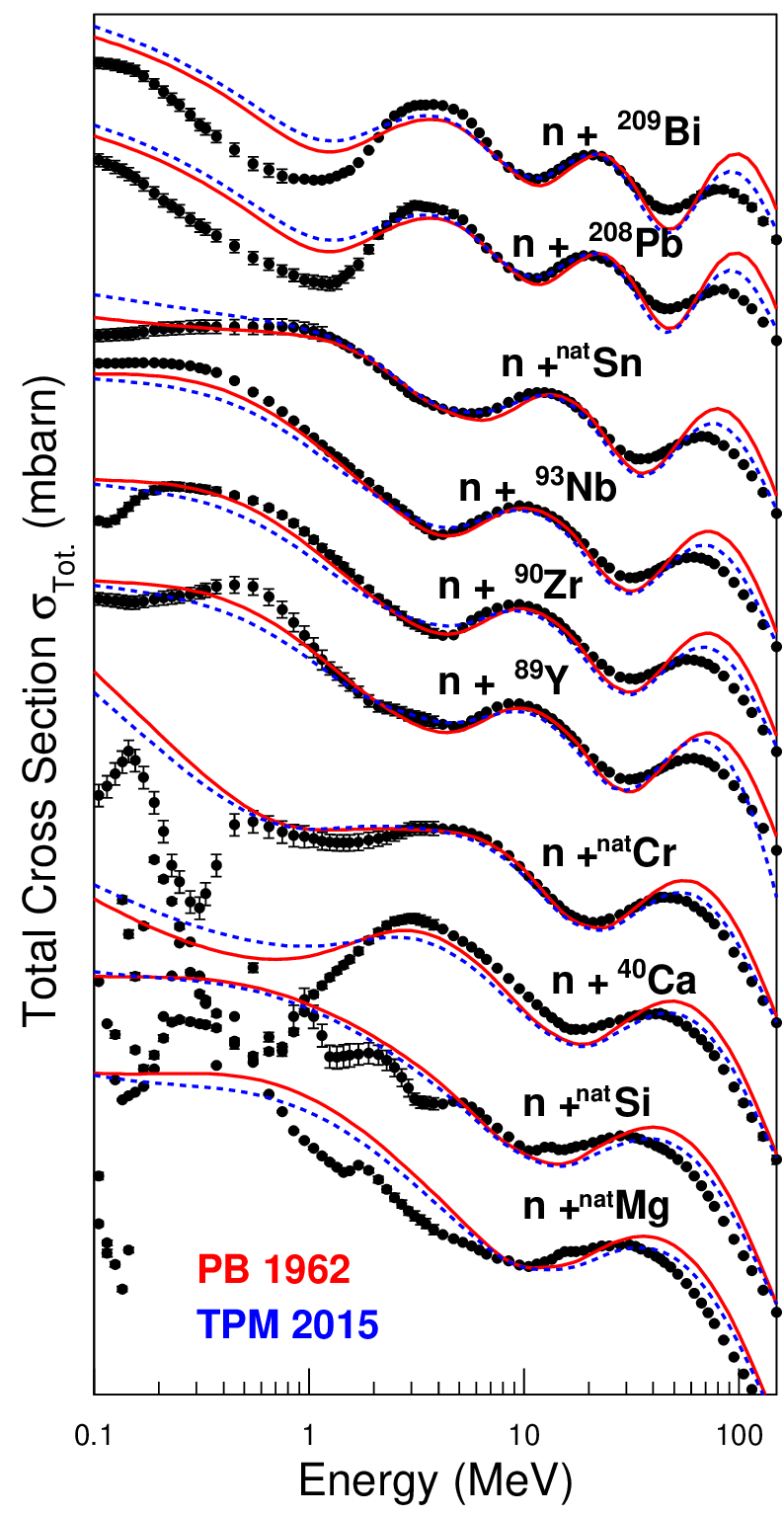}
  \caption{Total cross section for neutron-nucleus scattering as functions of the beam energy.
    The data (symbols) are from Ref.~\cite{exfor_14}. Red solid curves correspond
    to PB potential \cite{perey_62} and blue dashed curves to that of TPM \cite{tian_15}.}
\label{fig::Sec_totale_MG_Bi_PB_TPM}} 
\end{figure}

\begin{figure}
\centering{\includegraphics[scale=0.7]{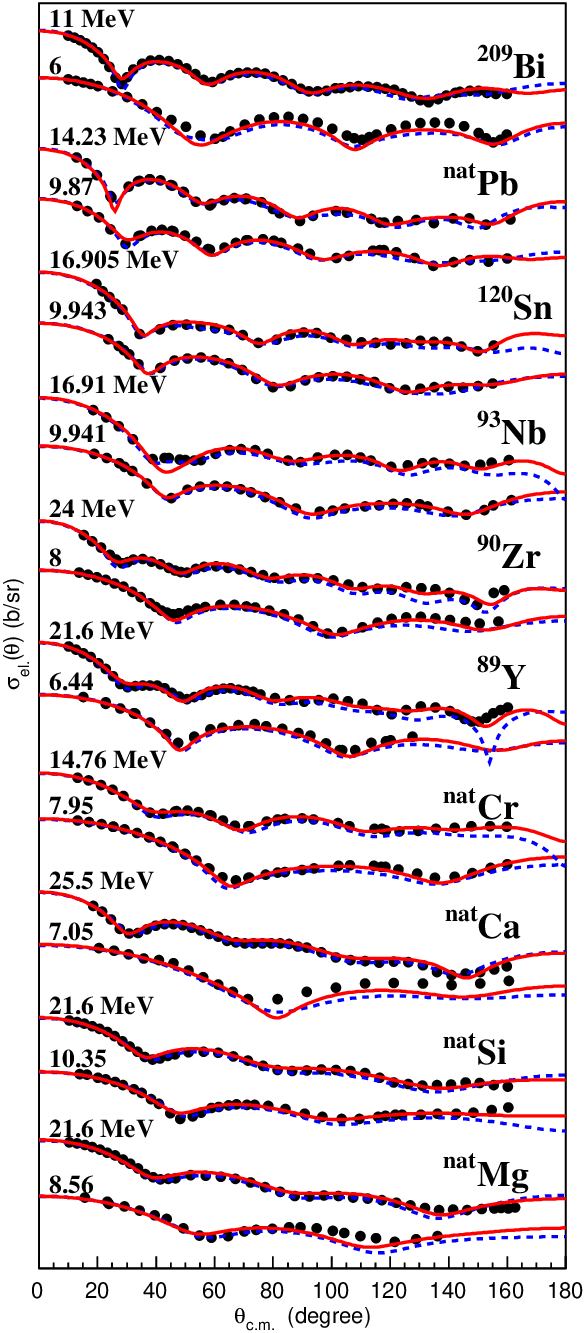}
  \caption{Differential cross sections for neutron elastic scattering as functions
    of the c.m. scattering angle. Red solid curves correspond to the PB
    potential \cite{perey_62} while blue dashed curves denote TPM \cite{tian_15} results.
    Data are denoted by symbols \cite{exfor_14}.}
\label{fig::Sec_elastique_Mg_Bi_PB_TPM}
}
\end{figure}

In order to illustrate the predictive power of PB potential,
in Fig.~\ref{fig::Sec_totale_MG_Bi_PB_TPM} we plot experimental and calculated cross sections
for ten target nuclei (Mg, Si, Ca, Cr, Y, Zr, Nb, Sn, Pb, Bi). Solid red and dashed blue curves
denote results from PB and TPM parametrizations, respectively. The experimental total cross
section has been averaged on the energy. The effect of this averaging is twofold. At low incident
energies it filters out the compound nucleus contribution \cite{feshbach_54}. It also allows to
attenuate discrepancies coming from merging different data sets. In
Fig.~\ref{fig::Sec_elastique_Mg_Bi_PB_TPM} we plot the experimental and calculated differential
cross sections as functions of the scattering angle in the center of mass reference frame.
For the sake of conciseness, we show only two angular distributions for each target nucleus.
Here we include results for PB and TPM parametrizations adopting the same conventions as in
Fig.~\ref{fig::Sec_totale_MG_Bi_PB_TPM}.

While angular distributions are very well reproduced, the agreement for the total cross
section is poor in the 0.4~-~28~MeV energy range for both parametrizations. It should be noted,
however, that when Perey and Buck developed their potential in the early sixties they did
not have the current wealth of measurements.
TPM potential provides a better agreement for high-energy total cross
sections. This is most likely due to the presence of a volume imaginary term in
their parametrization. Note, however, that the experimental total cross section is
systematically overestimated in the range 50-150~MeV. 

\section{Nonlocal and dispersive optical model}
\label{sec::nldisp}

Although neither microscopic nor \textit{ab-initio} approaches for nucleon-nucleus
scattering are currently suited for massive and precise data evaluation, they can
provide guidance to phenomenological models. On the other side, phenomenological
approaches can become very useful mainly due to their simplicity and flexibility.
The optical model potential is connected to microscopy through its correspondence with
the self-energy \cite{bell_59}. The challenge here is to incorporate microscopic
and \textit{ab-initio} features in the construction of phenomenological optical potentials. \\

One of these features is the nonlocality, emerging from Pauli exclusion principle,
in addition to dynamic polarization \cite{fraser_08}. In a recent work it has been
demonstrated that the nonlocality exhibits a bell-shape \cite{arellano_22}. These findings
emerge from microscopic descriptions of nucleon-nucleus collisions based on density-dependent
g-matrix folding model retaining nonlocalities at all stages \cite{arellano_22,arellano_21b}.
It is found that the nonlocality form factor $H$ (see Eq.~\eqref{eq::pbnl}) is weakly
energy-dependent. The leading energy dependence takes place in the strengths of the potential.
Moreover, the spin-orbit contribution results nonlocal. In the context of PB
approach \cite{perey_62}, this nonlocality is modeled through a Gaussian form factor.
We retain a single Gaussian nonlocality for real and imaginary components of the potential. 
As we will see, this is sufficient to achieve a reasonable description of experimental data.\\

Continuing the analogy with many-body physics, the optical potential can be separated into
an energy-independent Hartree-Fock and energy-dependent polarization terms. Then, causality
is ensured through a dispersion relation \cite{mahaux_86,mahzoon_th}. This dispersion relation
relates scattering and bound-state properties. Hence, it allows to determine scattering states
and bound states of the target-nucleus on the same footing using a single potential. 
A well-constrained phenomenological nonlocal dispersive potential contains in practice all the
correlations. Thus one can evaluate the quality of the microscopic description not only of
scattering observables but also other physical properties such as occupation
numbers, densities and bound state energies.\\

\subsection{Dispersion relation}

The radial form factor $U$ in Eq.~\eqref{eq::rad-ff} is now assumed energy dependent,
\begin{eqnarray}
  U(\tilde{r},E) = V(\tilde{r},E)+iW(\tilde{r},E), 
  \label{eq::U-Edep}
\end{eqnarray}
whereas the nonlocal form factor $H$ in Eq.~\eqref{eq::pbnl} is kept energy independent.
Now, a dispersion relation applies relating the energy-dependent real and imaginary
contributions in Eq.~\eqref{eq::U-Edep}, 
\begin{eqnarray}
  V(\tilde{r},E)&=&V_{0}(\tilde{r})+\Delta V(\tilde{r},E), \nonumber \\
 \Delta V(\tilde{r},E)&=& \frac{\cal P}{\pi}\int_{-\infty}^{+\infty}
 \frac{W(\tilde{r},E')}{E'-E} dE',
 \label{eq::disp}
\end{eqnarray}
where ${\cal P}$ denotes the principal value of the integral \cite{mahaux_91}.
$V_{0}$ is an energy-independent potential often referred to as Hartree-Fock, in analogy to
the static self-energy in many-body physics \cite{fetter_71}. When possible,
the dispersive term $\Delta V$ is obtained analytically, otherwise it is calculated
by numerical integration. We assume the energy dependence affects only the strengths
of the potential. The diffuseness, radius and nonlocality are assumed energy independent.

\subsection{Real potentials}

The real potential is assumed nonlocal with a Gaussian nonlocal form factor
$H$ as in Eq.~\eqref{eq::pbnl} and energy independent. Its radial form factor
reads
\begin{eqnarray}
 V_{0}(\tilde{r})&=&V_{V}^{\mathrm{NL}}f(\tilde{r},R,a)
 - V_{S}^{\mathrm{NL}} 4a\frac{df(\tilde{r},R,a)}{d\tilde{r}} \nonumber \\
 &-& 2 V_{so}^{\mathrm{NL}}\left(\frac{\hbar}{m_{\pi}c}\right)^{2}
 \frac{1}{\tilde{r}}\frac{df(\tilde{r},R,a)}{d\tilde{r}}\;
 \mathbf{l}.\mathbf{s} .
\end{eqnarray}
We include a real nonlocal surface potential $V_{S}^{\mathrm{NL}}$  that improves the agreement
with the elastic angular distribution data. The spin-orbit potential is nonlocal with
the same nonlocality than the central term. The dispersive component,
$\Delta V(\tilde{r},E)$, stemming from the energy-dependent imaginary potentials are added
as described in Eq.~\eqref{eq::disp}. 

\subsection{Imaginary potentials}

As observed in Fig.~\ref{fig::Sec_totale_MG_Bi_PB_TPM}, the addition of a volume
imaginary contribution, as proposed by TPM \cite{tian_15}, helps to better reproduce
total cross section above 50~MeV. Following  this prescription, we have tested several
values of nonlocality parameters $\beta$ for the volume imaginary potential. We have
also investigated different energy dependencies. From these studies we conclude that a
local volume imaginary potential is the best suited. The energy-dependency retained
for the volume and the surface terms are described below. \\
\indent The local imaginary volume contribution reads  
\begin{eqnarray}
 W^{\mathrm{L}}(r,E) &=& W_{V}^{\mathrm{L}}(E) f(r,R,a),
\end{eqnarray}
whereas the nonlocal imaginary surface contribution reads
\begin{eqnarray}
  W^{\mathrm{NL}}(\tilde{r},E)
     &=& - W_{S}^{\mathrm{NL}}(E) 4a\frac{df(\tilde{r},R,a)}{d\tilde{r}} \nonumber \\
  &-&  2 W_{so}^{\mathrm{NL}}(E)\left(\frac{\hbar}{m_{\pi}c}\right)^{2}
  \frac{1}{\tilde{r}}\frac{df(\tilde{r},R,a)}{d\tilde{r}}\;
 \mathbf{l}.\mathbf{s}. \nonumber \\
\end{eqnarray}
It should be noted that the imaginary spin-orbit is chosen to be nonlocal. 

\subsubsection{Surface imaginary depth}

For the depth of the nonlocal imaginary surface strength, we use a Brown-Rho
shape \cite{brown_81} modified by an exponential falloff,
\begin{eqnarray}
  W_{S}^{\mathrm{NL}}(E)=A_{S}^{\pm}(E-E_{F})^{n}
  \frac{\exp(-C_{S}\left|E-E_{F}\right|)}{(E-E_{F})^{n}+B_{S}^{n}}.
\end{eqnarray}
Its energy dependence is asymmetric with respect to the neutron Fermi energy
$E_{F}=-[S_{n}(Z,N)-S_{n}(Z,N+1)]/2$. There is a single depth value $A_{S}^{+}$
for $E>E_{F}$, and another one $A_{S}^{-}$ for $E<E_{F}$. We have investigated
different values for the parameter $n$, reaching better fits with $n=2$. We
also found that coefficients $A_{S}^{-}$, $B_{S}$ and $C_{S}$ exhibit a very weak
dependence on target-nucleus mass, so that it is reasonable to assume them
constant. In contrast, $A_{S}^{+}$ decreases linearly with respect to the mass $A$.

\subsubsection{Volume imaginary depth}

For energies around the Fermi energy, within the interval $[E_{F}-E_{V}^{-}$, $E_{F}+E_{V}^{+}]$,
the depth of the local imaginary volume component is of Brown-Rho type \cite{brown_81}
but with different depths $A_{V}^{\pm}$, depending on whether the projectile energy is
above ($A_{V}^{+}$) or below ($A_{V}^{-}$) the Fermi energy, namely
\begin{equation}
W_{V}^{\mathrm{L}}(E)=\frac{{\textstyle A_{V}^{\pm}(E-E_{F})^{2}}}{{\textstyle (E-E_{F})^{2}+B_{V}^{2}}}.
\end{equation}
For energies $E$ above $E_{F}+E_{V}^{+}$, the form proposed by Mahaux and Sartor \cite{mahaux_91}
is applied, thus 
\begin{eqnarray}
W_{V}^{\mathrm{L}}(E)&=&\frac{{\textstyle A_{V}^{+}(E-E_{F})^{2}}}{{\textstyle (E-E_{F})^{2}+B_{V}^{2}}} \nonumber \\
&+&\alpha\left[\sqrt{E}+\frac{(E_{F}+E_{V}^{+})^{3/2}}{2E}-\frac{3}{2}\sqrt{E_{F}+E_{V}^{+}}\right].\nonumber \\
\end{eqnarray}
In the case of $E$ below $E_{F}-E_{V}^{-}$, the strength is given by
\begin{eqnarray}
W_{V}^{\mathrm{L}}(E)&=&\frac{{\textstyle A_{V}^{-}(E-E_{F})^{2}}}{{\textstyle (E-E_{F})^{2}+B_{V}^{2}}} \nonumber\\
&\times& \left[1-\frac{\left(E-E_{F}+E_{V}^{-}\right)^{2}}{\left(E-E_{F}+E_{V}^{-}\right)^{2}+\left(E_{V}^{-}\right)^{2}}\right].
\end{eqnarray}

\subsubsection{Spin-orbit imaginary depth}

For the spin orbit term, the imaginary surface strength is taken symmetrical with
respect to the Fermi energy. It reads
\begin{eqnarray}
  W_{so}^{\mathrm{NL}}(E)=\frac{A_{so}(E-E_{F})^{2}}{(E-E_{F})^{2}+C_{so}^{2}}
  -\frac{B_{so}(E-E_{F})^{2}}{(E-E_{F})^{2}+D_{so}^{2}},\nonumber \\
\end{eqnarray}
consisting of the difference of two Brown-Rho forms, allowing an analytical
expression for the dispersive term $\Delta V_{so}^{\mathrm{NL}}(E)$ \cite{vanderkam_00}.

\section{Parameter calibration}
\label{sec::opt}

Once the potential form factors are specified, we look for the parameters that best
reproduce the experimental data. For each nucleus we carry out an iterative parameter
calibration to minimize the difference between data and the nonlocal dispersive optical
model results. Then one can identify the dependence of the parameters on the nucleus
mass $A$ and the energy. The variation of the parameters with respect to the mass and the
energy is expected to be smooth, helping the model to gain predictive power. This
parameter search is highly time-consuming, being necessary the use of parallel computing.
For each nucleus, over 2000 processors are used for twenty-four hours to find the parameters. 
In addition, for low neutron energy (below about 15~MeV), compound neutron emission is
described with Hauser-Feshbach formalism \cite{hauser_52} using NLD transmission coefficients
in TALYS code \cite{koning_08}. This process is accounted for in the calibration process. 
\\

Our search procedure makes extensive use of the comprehensive EXFOR
database \cite{exfor_14}. In the energy range from 1~keV to 250~MeV, the
experimental data of the following twenty-two elements are used to
determine our parameters: C, O, Mg, Al, Si, S, Ca, Ti, Cr, Fe, Ni, Cu,
Y, Zr, Nb, Mo, Sn, Ce, Au, Hg, Pb, Bi. About 100~energy points are
needed to describe accurately the variations of the total cross section
in the energy range 1~keV to 250~MeV. As already mentioned, for neutron
energies below a few~MeV and especially for light nuclei, the experimental
total cross section must be averaged because of the presence of compound
nucleus resonances. This averaged total cross section (see
Fig.~\ref{fig::Sec_totale_MG_Bi_PB_TPM}) is retained throughout the calibration.
After each step of parameter search, we calculate bound state energies
as well as analyzing powers.
\\

\begin{table}[tbph]
\begin{centering}
\begin{tabular}{ll}
\hline 
\hline 
\multicolumn{2}{c}{Nonlocal real depth}\tabularnewline
\hline
 & \tabularnewline
$V_{V}^{\mathrm{NL}}$ (MeV)  & $-69.71-1.140 \times 10^{-2}$ A \tabularnewline
$V_{S}^{\mathrm{NL}}$ (MeV)  & $-8.600-8.000 \times 10^{-3}$ A \tabularnewline    
$V_{so}^{\mathrm{NL}}$ (MeV)  & $-9.787-1.140 \times 10^{-2}$ A\tabularnewline    
 & \tabularnewline
\hline 
\multicolumn{2}{c}{Nonlocal surface imaginary depth}\tabularnewline
\hline 
 & \tabularnewline
\multicolumn{2}{l}{$W_{S}^{\mathrm{NL}}(E)=A_{S}^{\pm}(E-E_{F})^{2}\frac{{\textstyle \exp(-C_{S}\left|E-E_{F}\right|)}}{{\textstyle (E-E_{F})^{2}+B_{S}^{2}}}$}\tabularnewline
\multicolumn{2}{c}{}\tabularnewline
$A_{S}^{+}$ (MeV)  & $-19.62-1.500 \times 10^{-2}$A,\hspace{5mm} for $E_{F}<E$\tabularnewline
$A_{S}^{-}$ (MeV)  & $-16.00$,\hspace{5mm} for $E<E_{F}$\tabularnewline
$B_{S}$ (MeV)  & $11.11$ \tabularnewline
$C_{S}$ (MeV$^{-1}$)  & $9.200 \times 10^{-3}$ \tabularnewline
 & \tabularnewline
\hline 
\multicolumn{2}{c}{Local volume imaginary depth}\tabularnewline
\hline 
 & \tabularnewline
\multicolumn{2}{l}{\underline{For $E>E_{F}+E_{V}^{+}:$}}\tabularnewline
& \tabularnewline
\multicolumn{2}{l}{$W_{V}^{\mathrm{L}}(E)=\frac{{\textstyle A_{V}^{+}(E-E_{F})^{2}}}{{\textstyle (E-E_{F})^{2}+B_{V}^{2}}}$}\tabularnewline
\multicolumn{2}{c}{$+\alpha\left(\sqrt{E}+\frac{(E_{F}+E_{V}^{+})^{3/2}}{2E}-\frac{3}{2}\sqrt{E_{F}+E_{V}^{+}}\right).$}\tabularnewline
 & \tabularnewline
\multicolumn{2}{l}{\underline{For $E_{F}-E_{V}^{-}<E<E_{F}+E_{V}^{+}:$}}\tabularnewline
 & \tabularnewline
\multicolumn{2}{l}{$W_{V}^{\mathrm{L}}(E)=\frac{{\textstyle A_{V}^{\pm}(E-E_{F})^{2}}}{{\textstyle (E-E_{F})^{2}+B_{V}^{2}}}.$}\tabularnewline
 & \tabularnewline
\multicolumn{2}{l}{\underline{For $E<E_{F}-E_{V}^{-}$:}}\tabularnewline
 & \tabularnewline
\multicolumn{2}{l}{$W_{V}^{\mathrm{L}}(E)=\frac{{\textstyle A_{V}^{-}(E-E_{F})^{2}}}{{\textstyle (E-E_{F})^{2}+B_{V}^{2}}}$}\tabularnewline
\multicolumn{2}{c}{$\times\left(1-\frac{\left(E-E_{F}+E_{V}^{-}\right)^{2}}{\left(E-E_{F}+E_{V}^{-}\right)^{2}+\left(E_{V}^{-}\right)^{2}}\right),$}\tabularnewline
 & \tabularnewline
 with\tabularnewline
$A_{V}^{+}$ (MeV)  & $-32.40-2.000 \times 10^{-2}$A,\ \ for $E>E_{F}$\tabularnewline
$A_{V}^{-}$ (MeV)  & $-8.400$A,\ \ for $E<E_{F}$ \tabularnewline
$B_{V}$ (MeV)     & $ 135.0$ \tabularnewline
$E_{V}^{+}$ (MeV)  & $40.00-9.000 \times 10^{-2}$A \tabularnewline
$E_{V}^{-}$ (MeV)  & $25.50$ \tabularnewline
$\alpha$ (MeV$^{1/2}$)  & $3.000 \times 10^{-1}+2.000 \times 10^{-3}$A\tabularnewline
 & \tabularnewline
\hline 
\multicolumn{2}{c}{Nonlocal spin-orbit imaginary depth}\tabularnewline
\hline 
 & \tabularnewline
\multicolumn{2}{c}{$W_{so}^{\mathrm{NL}}(E)=\frac{\textstyle A_{so}(E-E_{F})^{2}}{\textstyle (E-E_{F})^{2}+C_{so}^{2}}-\frac{\textstyle B_{so}(E-E_{F})^{2}}{\textstyle (E-E_{F})^{2}+D_{so}^{2}}$}\tabularnewline
 & \tabularnewline
$A_{so}$ (MeV)  & 4.893 \tabularnewline
$B_{so}$ (MeV)  & 2.447 \tabularnewline
$C_{so}$ (MeV)  & 50.00 \tabularnewline
$D_{so}$ (MeV)  & 3.900 \tabularnewline
\hline
\hline
\end{tabular}
\par\end{centering}
\caption{Energy and mass dependency of real and imaginary potential strengths.}
\label{tab::parametres_potentiel_imaginaire_pbevol} 
\end{table}
\begin{table}[tbph]
\begin{centering}
\begin{tabular}{ll}
\hline 
\hline 
\multicolumn{2}{c}{Geometrical parameters}\tabularnewline
\hline 
$r_{0}$ (70 $<$ A) (fm)  & $1.1446+2.4200\times10^{-4}$ A \tabularnewline
$r_{0}$ (A {$<$} 70) (fm)  & $9.4860 \times 10^{-1}+8.8000\times10^{-3}\mathrm{A}$ \tabularnewline
 & $-1.3200\times10^{-4}\mathrm{A}^{2}+7.1000\times10^{-7}\mathrm{A}^{3}$\tabularnewline
$a$ (fm)  & $6.1600 \times 10^{-1}-1.8200\times10^{-4}$ A \tabularnewline
$\beta$ (fm)  & 0.915 \tabularnewline
\hline
\hline 
\end{tabular}
\par\end{centering}
\caption{Nonlocality and mass dependence of depth, radii and diffuseness of NLD potential.}
\label{tab::parametres_potentiel_pbevol_reel_geometrie} 
\end{table}

\begin{figure}
  \centering{\includegraphics[scale=0.9]{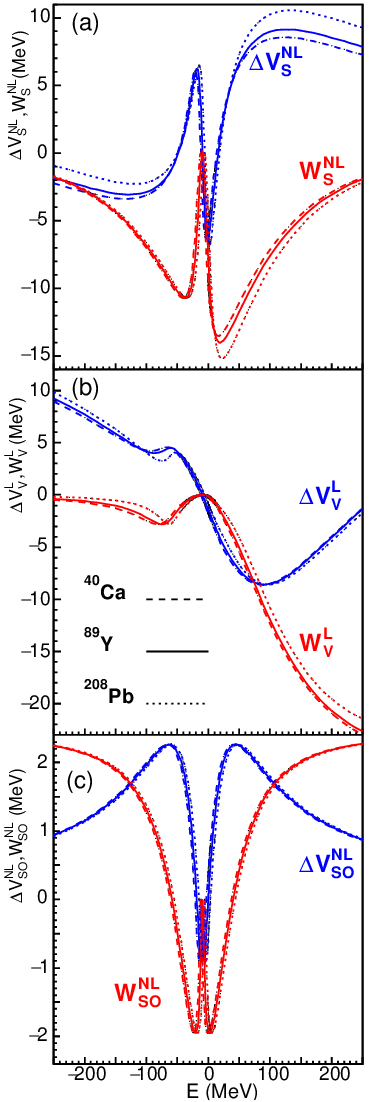}
  \caption{Depths $W$ (red curve) and dispersive contribution $\Delta V$
      of surface (a), volume (b) and spin-orbit (c) imaginary potentials
      as functions of the energy. Dashed, solid and dotted curves denote
      results for $^{40}$Ca, $^{89}$Y and $^{208}$Pb, respectively. 
      \label{fig::variations_profondeurs_potentiel}}}
\end{figure}

The resulting parameters for strengths of the real and imaginary potentials
are summarized in Table~\ref{tab::parametres_potentiel_imaginaire_pbevol}. 
In Fig.~\ref{fig::variations_profondeurs_potentiel}, we present the strengths
of the surface [panel (a)], volume [panel (b)] and spin-orbit [panel (c)] imaginary
contributions (red curves) together with their respective dispersive contributions
(blue curves) for three target-nuclei: $^{40}\mathrm{Ca}$, $^{89}\mathrm{Y}$ and
$^{208}\mathrm{Pb}$. In panel (a), $W_{S}^{\mathrm{NL}}$ is non symmetrical with
respect to the Fermi energy. This asymmetry results from the fact that the phase
space of particle levels for
$E \ll E_{F}$ is significantly larger than that of hole levels for $E \gg E_{F}$.
Therefore the contributions from
two-particle–one-hole states for $E \ll E_{F}$ will be larger than that for
two-hole–one-particle states at $E \gg E_{F}$ \cite{charity_07}.
In panel (c), $W_{so}^{\mathrm{NL}}$ is independent of the mass and symmetrical
with respect to the Fermi energy. 
The real depths of the volume surface and spin-orbit potential
($V_{V}^{\mathrm{NL}}$, $V_{S}^{\mathrm{NL}}$ and $V_{so}^{\mathrm{NL}}$)
are independent of the energy of the projectile. The variations as a function
of the mass of the nuclei are weak for the surface and volume depths, as shown in
Table~\ref{tab::parametres_potentiel_imaginaire_pbevol}.

\begin{figure}
  \centering{\includegraphics[scale=0.45]{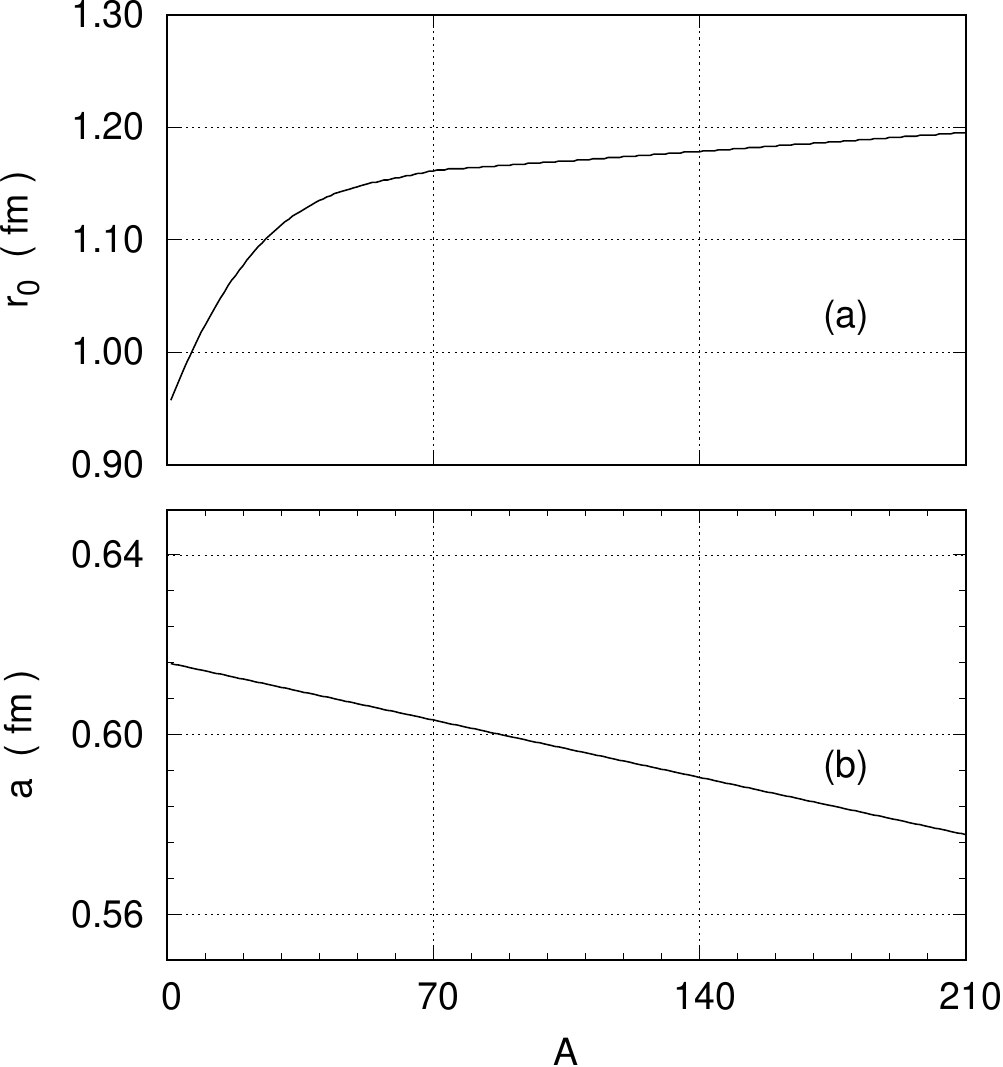}
    \caption{Reduced radius $r_{0}$ (a) and diffuseness $a$ (b) as functions of the
      target mass $A$ for the nonlocal dispersive model.
      \label{fig::variations_r_a}}}
\end{figure}

The optimum nonlocality range $\beta$ is found to be 0.915~fm, larger than in previous works
\cite{perey_62,tian_15}. The same radius and diffuseness are used for real and
imaginary contributions. In Fig.~\ref{fig::variations_r_a} we plot the reduced radius
$r_{0}$ [panel (a)] and
the diffuseness $a$ [panel (b)] as functions of the mass $A$. The reduced radius
increases with mass in contrast to diffuseness, which is a linear function decreasing
as the mass increases. For the former, the variation is linear for masses greater
than 70, while for smaller masses the expression of the radius behaves as a polynomial
of degree 3. Results are summarized in
Table~\ref{tab::parametres_potentiel_pbevol_reel_geometrie}.

\section{Comparison between local and nonlocal dispersive models}
\label{sec::comp_local_nl}

We examine differences in the calculated scattering observables for
neutron scattering as implied by the local dispersive (LD) model
of Ref.~\cite{morillon_07} and the nonlocal dispersive (NLD) model
presented in this work.

\begin{figure*}
	\centering{
                \includegraphics[scale=0.7]{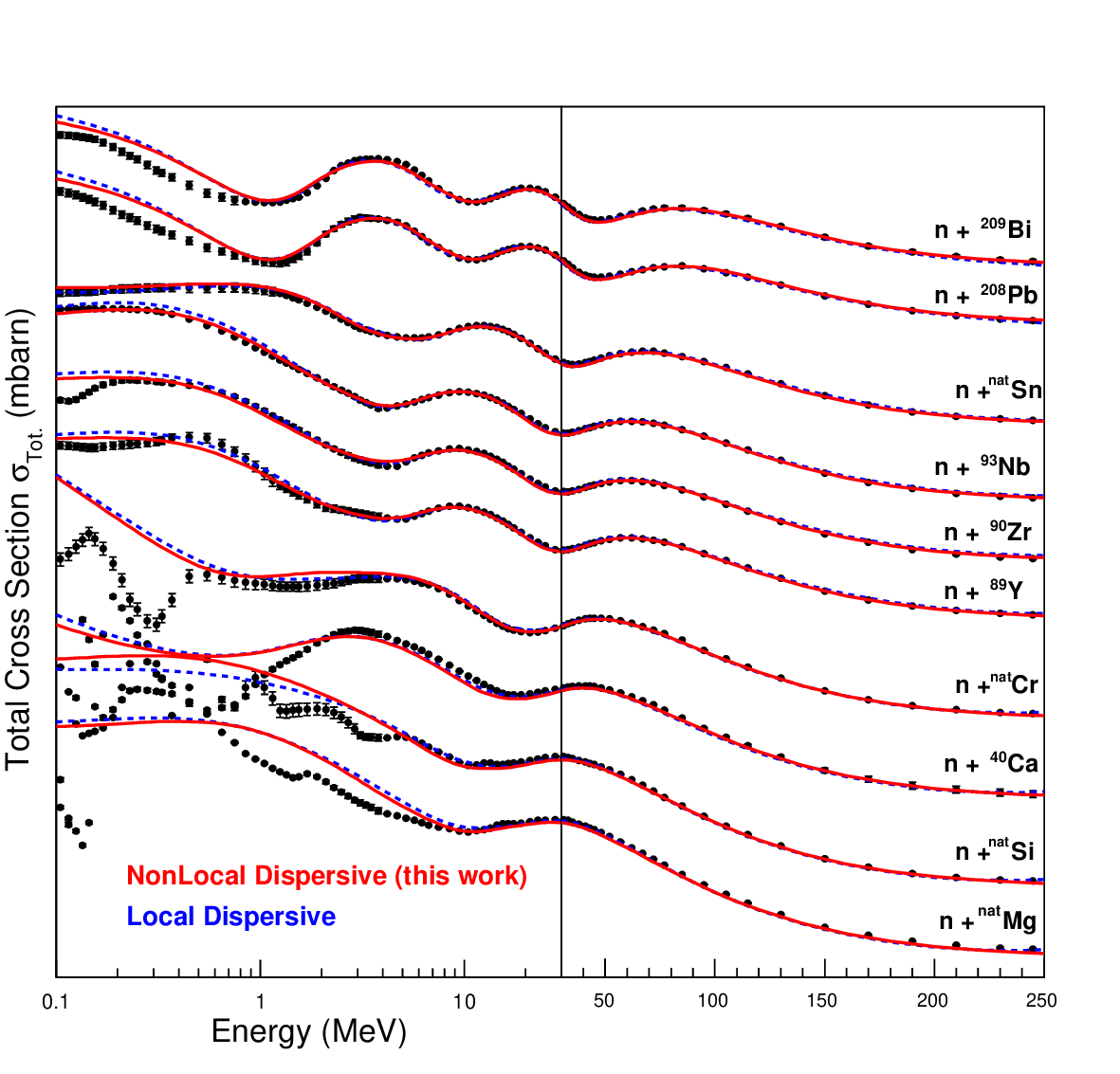}
		\caption{Total cross section for neutron-nucleus scattering as
                  function of the energy. The data (symbols) are from Ref.~\cite{exfor_14}. 
                  Red solid curves correspond to NLD potential (this work) and blue dashed
                  curves to LD potential \cite{morillon_07}.}
		\label{fig::Sec_totale_Pb_Zr_Ca_BMR_MR}
	} 
\end{figure*}

In Fig.~\ref{fig::Sec_totale_Pb_Zr_Ca_BMR_MR} we plot the total cross sections
for neutron-nucleus scattering as function of the energy. We include
ten targets with masses between 24~and~209. Dashed blue curves denote results from
LD model whereas solid red curves denote NLD results (this work). We
observe that both models yield comparable results for $\sigma_{T}$ for neutron
energies above $\sim\!10$~MeV. Differences are evidenced at the lower energies,
although their agreement with the data is comparable.

\begin{figure}
	\centering{
          \includegraphics[scale=0.7]{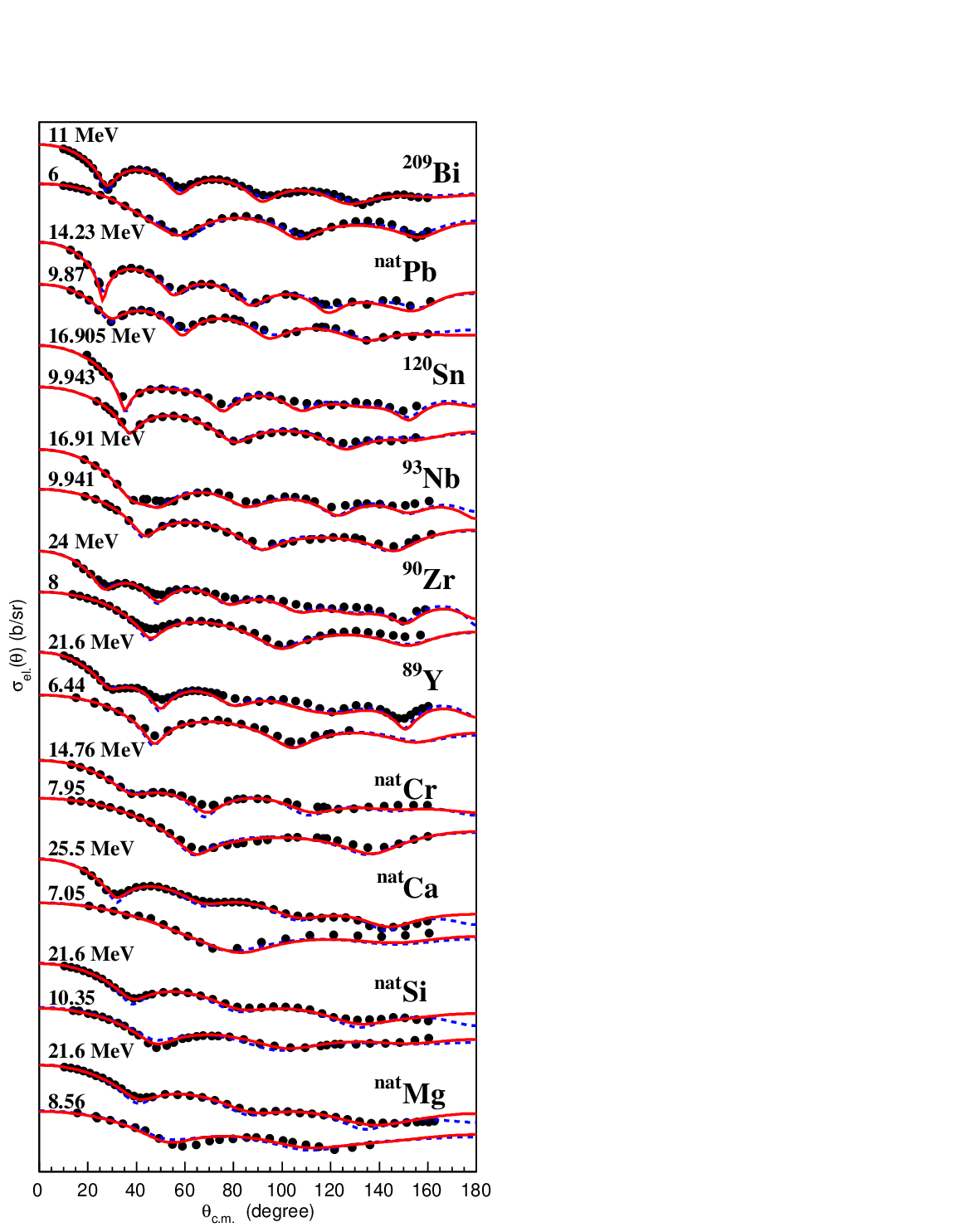}
		\caption{Differential cross sections for neutron elastic scattering. 
                  Data are denoted by symbols \cite{exfor_14}. Red solid curves
                  correspond to NLD potential (this work) and blue dashed
                  curves to LD potential \cite{morillon_07}.}
		\label{fig::Sec_elastique_BMR_MR}
	} 
\end{figure}

In Fig.~\ref{fig::Sec_elastique_BMR_MR} we show results for the differential
cross section as function of the scattering angle in the center-of-mass
reference frame ($\theta_{c.m.}$). The targets we include in this case are
the same as in Fig.~\ref{fig::Sec_totale_Pb_Zr_Ca_BMR_MR}, with neutron energies
ranging from 6 up to 25.5~MeV. Although the angular description of the data
appears comparable from both NLD and LD models, some differences are evidenced
in their maxima and minima, with the NLD approach in closer agreement with
the data.

\begin{figure}
  \centering{\includegraphics[scale=0.55]{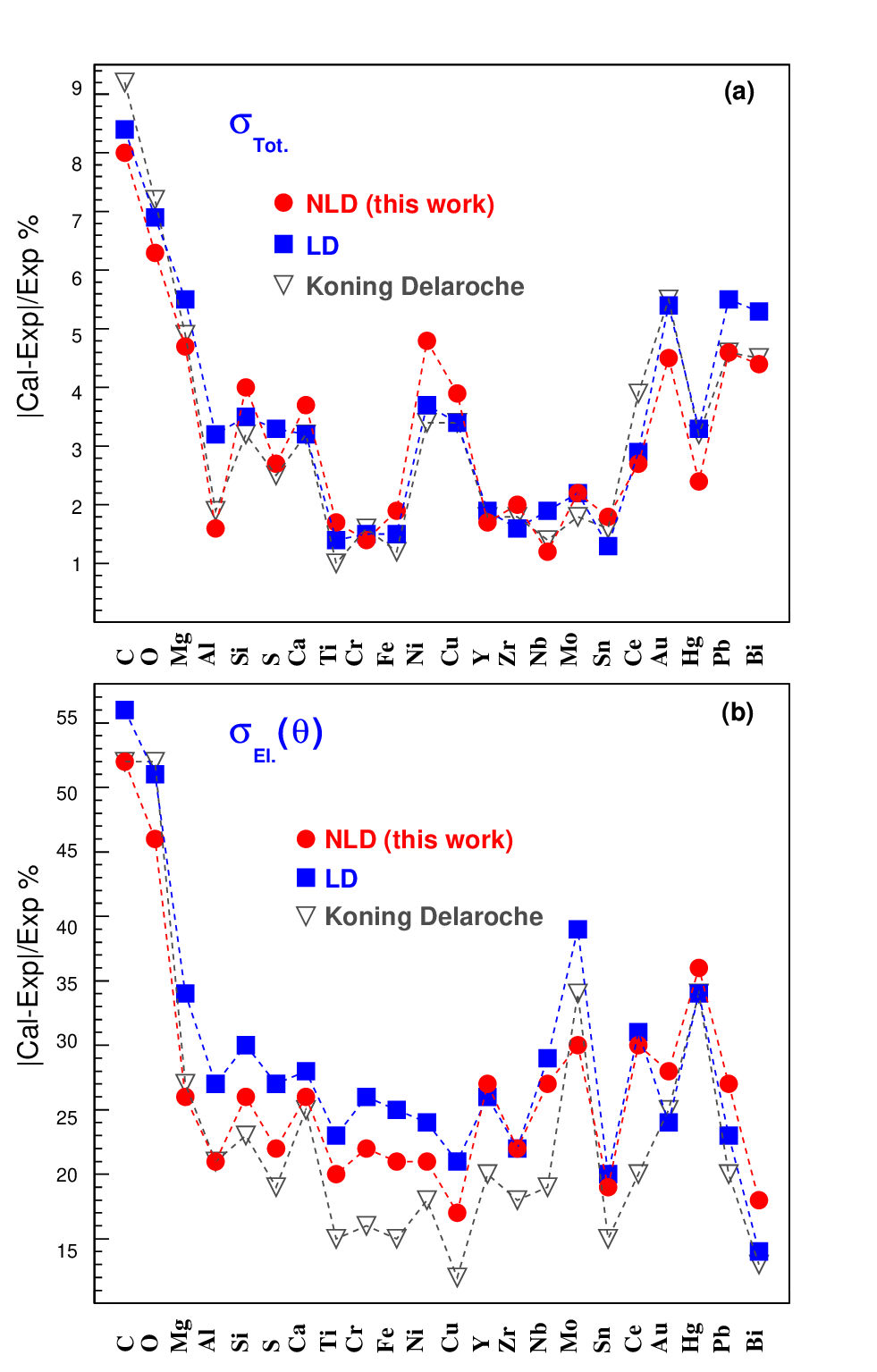}
    \caption{Relative differences experimental $vs.$ calculation expressed as percentages for each
      nucleus (dashed curves) for total cross section [panel (a)] and differential cross section
      [panel (b)]. Results are for NLD potentials (red circles), LD potential \cite{morillon_07}
      (blue square) and KD potential \cite{koning_03} (black triangle).}
    \label{fig::ecart_totale_elastique_KD_MR_BMR}}
\end{figure}

To compare the agreement with the data of NLD and LD approaches,
we have calculated relative differences with the data for all the cases
considered in the fit procedure. Results for these differences are
shown in Fig.~\ref{fig::ecart_totale_elastique_KD_MR_BMR}, where panel (a) summarizes
results for the total cross section, whereas panel (b) those for
the differential cross  section. Red circles (blue squares) denote
results from NLD (LD) approaches. Additionally, as reference, we
include in this analysis the KD non-dispersive model
(black triangles). From panel (a) we observe that all three models
yield comparable agreement in $\sigma_{T}$, of the order of 3\% for
medium-mass nuclei. However, for light nuclei the NLD model yields better
agreement (by roughly $\sim\!1$~\%) relative to the LD approach.
In the case of the differential cross section, panel (b) of 
Fig.~\ref{fig::ecart_totale_elastique_KD_MR_BMR} that the NLD approach is
in better agreement with the data relative to the LD model,
particularly for all targets with nuclear masses below 64.


\begin{figure}
   \centering{\includegraphics[scale=0.54]{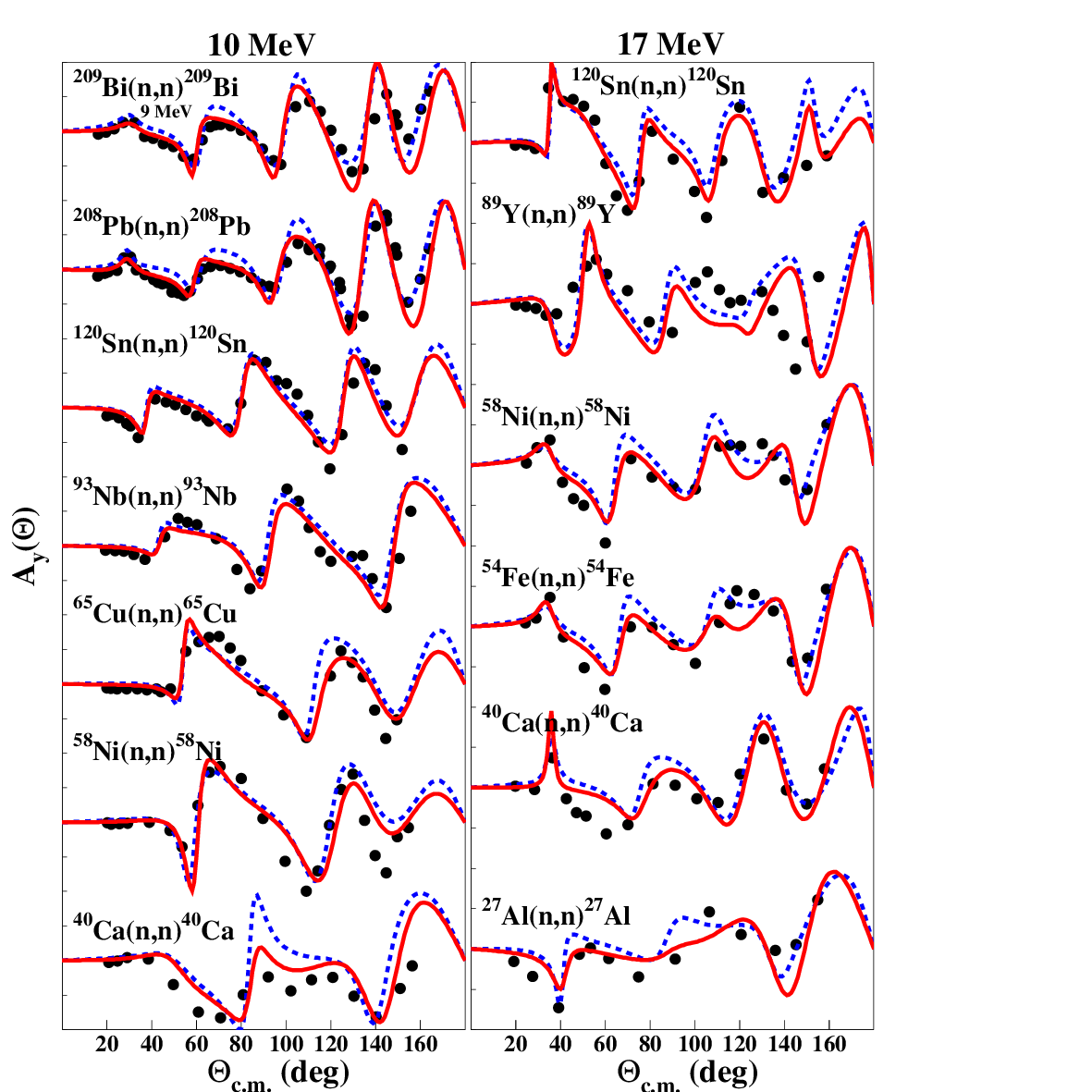}
     \caption{Analyzing powers for neutron scattering at 10 and 17~MeV. Data are denoted
       by symbols. Red solid curves are obtained with NLD potential and the blue dashed
       curves with LD potential.}
     \label{fig::pola_BMR_MR}} 
\end{figure}

In Fig.~\ref{fig::pola_BMR_MR} we show the analyzing powers (red curves) obtained from
our nonlocal OMP for two incident neutron energies (10~MeV and 17~MeV) with
the following nuclei: $^{27}$Al, $^{40}$Ca, $^{54}$Fe, $^{58}$Ni, $^{65}$Cu,  $^{89}$Y, $^{93}$Nb,
$^{120}$Sn, $^{208}$Pb and $^{209}$Bi. These results are compared with those calculated
with LD potential (blue dashed curves). This figure shows that the nonlocal
potential better reproduces the analyzing powers, particularly its maxima and minima. 

\begin{figure*}[tp]
  \centering{\includegraphics[scale=0.70]{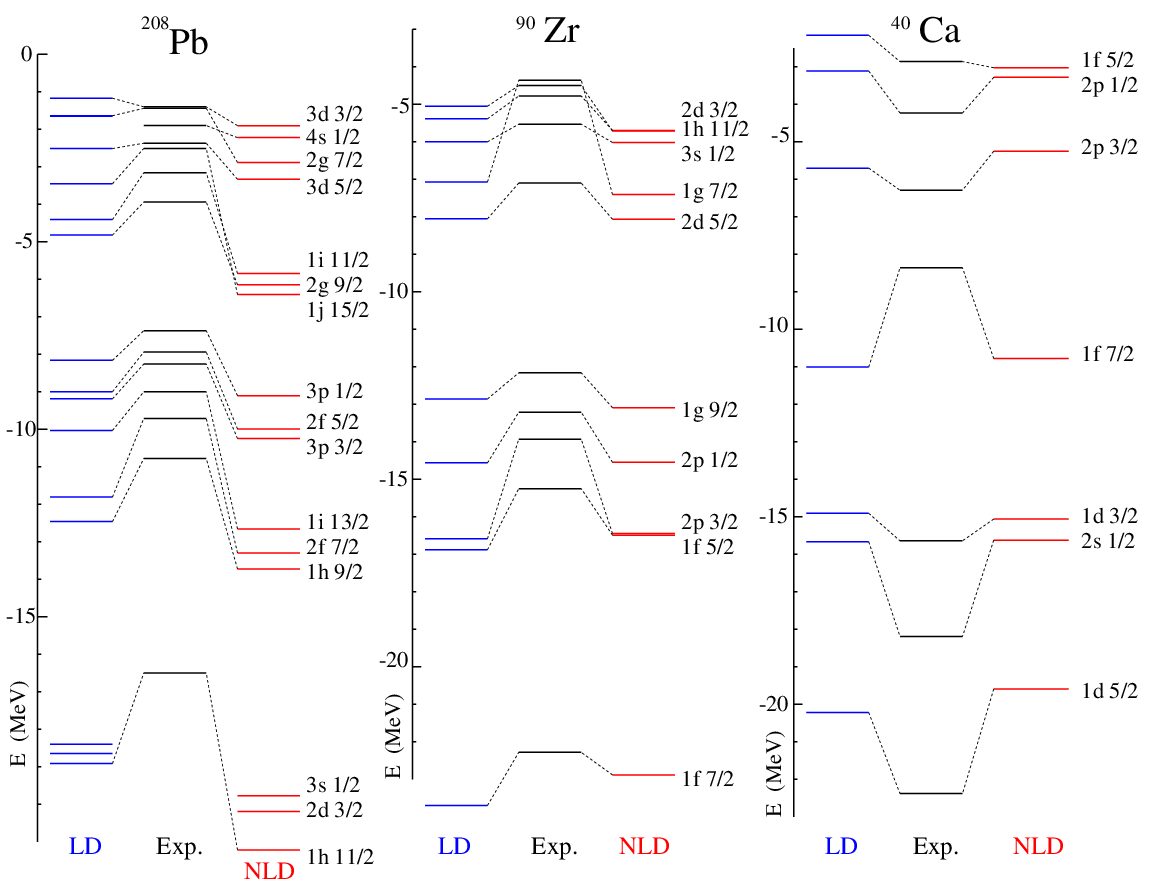}
    \caption{Neutron single-particle energies for $^{208}$Pb, $^{90}$Zr and $^{40}$Ca
             determined with LD, NLD potentials and from experiment. 
	    \label{fig::pb208zr90ca40_n}}} 
\end{figure*}

In Fig.~\ref{fig::pb208zr90ca40_n} we show the experimental and calculated neutron
single-particle energies for $^{208}$Pb, $^{90}$Zr and $^{40}$Ca. We include results
from LD and NLD global potentials. We observe comparable agreement of these two
models with the data, leaving room for improvement. 

\section{Uncertainties}
\label{sec::uncertainties}

\begin{figure}
	\centering{
		\includegraphics[scale=0.4]{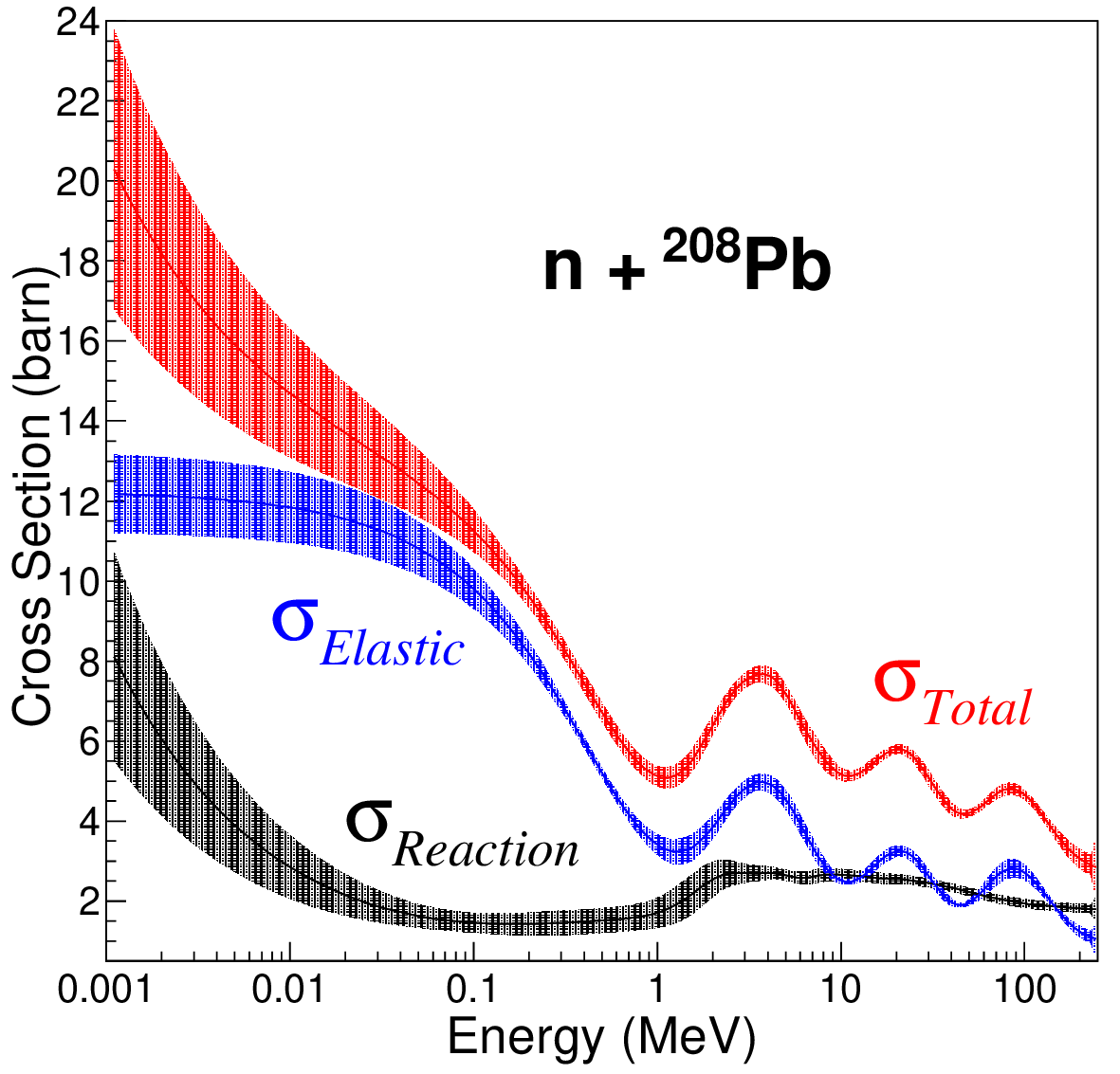}
		\caption{Uncertainties ($\pm 1 \sigma$) in the total (red), reaction (black)
                         and elastic (blue) cross sections for neutron scattering of $^{208}$Pb.}
		\label{fig::incertitudes}
	} 
\end{figure}

To assess the effect of random variations of the parameters of the NLD model on scattering
observables, we have calculated the standard deviation for the calculated cross sections
when geometrical parameters and strengths are varied. Specifically, in
Fig.~\ref{fig::incertitudes} we show the calculated elastic $\sigma_E$ (blue curves),
reaction $\sigma_R$ (black curves) and total $\sigma_T$ (red curves) cross sections for
n~+~$^{208}$Pb elastic scattering between 1~keV and 250~MeV. In this case, at each energy
we allowed uniform variations for radius (1\%), diffuseness (4\%), real volume and
surface depth (2\%) and range of nonlocality (2\%). Additionally, we allow for 20\%
variations of the depth of the imaginary terms. Thus, we performed stratified samplings
of parameters over a uniform distribution, resulting in two thousand sets of parameters
$(r, a, V_{S}, V_{V}, \beta, W_{S}, W_{V})$, each of them leading to a value for $\sigma_E$,
$\sigma_R$ and $\sigma_T$. In Fig.~\ref{fig::incertitudes} we plot the mean value (solid curves)
together with its $\pm 1 \sigma$ variations (shaded areas). It is interesting to note
that for energies above 0.2~MeV, the cross sections suffer rather moderate variations
under variations of parameters. Therefore, the limitations of the model shown in
Fig.~\ref{fig::Sec_totale_Pb_Zr_Ca_BMR_MR} for $\sigma_T$ cannot be attributed to the
parameters but to the assumed structure in the separable construction. This is an
indication that a more complete model is needed.

\section{Conclusions}
\label{sec::conc}

We have presented the first global, nonlocal and dispersive optical model potential
for neutron elastic scattering off spherical nuclei. The model is suited for beam energies  
up to 250~MeV and for target masses $16\le A \le 209$. Both central and spin-orbit terms
include a nonlocal Gaussian form factor of Perey-Buck type. Additionally, the imaginary volume
potential includes a local term. The strengths of the model are constrained by dispersion
relations, allowing a good description of integrated and angular scattering data.
The dispersion relation allows to determine scattering states and bound states of the
target-nucleus with a single potential. We are fully aware that there remains
room for improvement for the description of bound state properties. We obtain a better
description of the angular scattering data for $A \lesssim 65$ relative to the
LD approach \cite{morillon_07}.
These improvements appear more evident in the analyzing power, suggesting the relevance
of nonlocality in the spin-orbit term. This finding goes in line with microscopic studies
of nonlocality \cite{arellano_22}, where both central and spin-orbit components are
nonlocal. 
As already mentioned, an important advantage of global phenomenological approaches for nucleon
scattering lies on their simplicity, making it feasible computations which would require physical
information on processes involving the whole nuclear chart and over a broad energy range.
The challenge in this approach is to identify strengths and form factors as implied from
more fundamental approaches such as microscopic \cite{arellano_89,crespo_90,elster_90,vorabbi_16}
or \textit{ab-initio} \cite{hagen_12,rotureau_17,idini_19}. Efforts along this line are under way. 
 
\begin{acknowledgements}
  This work was performed using HPC resources from CCRT.
  H.F.A. thanks the hospitality of colleagues of CEA,DAM,DIF during his stay at Bruy\`eres-le-Ch\^atel.
\end{acknowledgements}

\appendix

\section{Resolution of the integro-differential Schr\"{o}dinger equation}
\label{sec::wf-calc}


The search procedure for strengths and geometrical parameters implies comparisons between
data and scattering observables resulting from the nonlocal optical model we have discussed.
This requires solving the integro-differential Schrödinger equation in Eq.~\eqref{eq::intdiff}
for several targets, energies and varying parameters. Thus, we need a fast and accurate
method to obtain the scattering wave functions. Along this line, Eq.~\eqref{eq::intdiff}
is solved by expanding $u_{jl}$ on the Chebyshev polynomial basis. Since these polynomials are
defined in the interval {[}-1,+1{]} we use the new variable $x=\frac{2}{R_{M}}r-1$. The quantity
$R_{M}$ represents the maximum value of the variable $r$ (it defines the radius where the potential
becomes negligible). Thus the radial wave function $u$ is written as a linear combination of
$N$ Chebyshev polynomials $T_{k}$ weighted by the coefficients $C_{k}$. It reads
\begin{eqnarray}
  u\left[\frac{R_{M}}{2}(x+1)\right]
  \simeq
  \sum_{k=0}^{N-1}\overset{/}{\vphantom{a}}C_{k}T_{k}(x),\quad x\in\lbrack-1,+1],
\end{eqnarray}
where the symbol $\overset{/}{}$ indicates that the first term of the  
sum is divided by a factor of 2, following the notation of Masson and
Handscomb \cite{masson_03}.

To obtain the radial part of the wave function we need to find the coefficients $C_{k}$
with $k=0,\ldots,N-1.$ That is, we need $N$ equations. The boundary conditions
at $r=0$, and $r=R_{M}$, yield two equations. Then considering that $T_{k}(1)=1$, and
$T_{k}(-1)=\left(-1\right)^{k}$, we get
\begin{eqnarray}
u(r=0) &=& \sum_{k=0}^{N-1}\overset{/}{\vphantom{a}}(-1)^{k}C_{k}=0,\label{eq_lo_intdiff_conditions_limites-1}  \\
u(r=R_{M}) &=& \sum_{k=0}^{N-1}\overset{/}{\vphantom{a}}C_{k}=1+i,
\end{eqnarray}
where $1+i$, is an arbitrary boundary value with no impact on the calculated phaseshift
\cite{joachain_75}. It thus remains to find $N-2$ equations in order to determine the
$C_{k}$ coefficients. This is done by evaluating Eq.~\eqref{eq::intdiff} at each of the $N-2$
roots $x_{n}$ of the $T_{N-2}$ Chebyshev polynomial. These roots are
\begin{equation}
  x_{n}=\cos\frac{(n-1/2)\pi}{N-2}\quad n=1,...,N-2.
  \label{eq:nm2_zero_cheby}
\end{equation}
In the case of the second derivative of the radial part of the wave function,
we take advantage of the second derivative of the Chebyshev polynomials,
\begin{equation}
  \left.\frac{d^{2}u\left(\frac{R_{M}}{2}(x+1)\right)}{dx^{2}}\right\vert _{x=x_{n}}\hspace{-0.25cm}=\hspace{-0.2cm}\sum_{k=2}^{N-1}\hspace{-0.1cm}C_{k}\hspace{-0.25cm}\sum_{\substack{m=0\\
m-r\ \mathrm{even}
}
}^{k-2}\hspace{-0.25cm}\overset{/}{\vphantom{a}}\negthickspace(k-m)k(k+m)T_{m}(x_{n}).\label{eq:derivee_seconde_chebyshev}
\end{equation}
In the case of a local potential, where we denote
\begin{eqnarray}
\mathcal{V^{\mathrm{L}}}(r)=\frac{l(l+1)}{r^{2}}+\frac{2\mu}{\hbar^{2}}V(r),
\end{eqnarray}
the $N-2$ equations representing Eq.~\eqref{eq::intdiff} become
\begin{eqnarray}
\frac{4}{R_{M}^{2}}\sum_{k=2}^{N-1}C_{k}\sum_{\substack{m=0\\
k-m\ \mathrm{even}
}
}^{k-2}\overset{/}{\vphantom{a}}(k-m)k(k+m)T_{m}(x_{n}) \nonumber \\
+\sum_{k=0}^{N-1}\overset{/}{\vphantom{a}}C_{k}T_{k}(x_{n})\left(\frac{2\mu E}{\hbar^{2}}-\mathcal{V^{\mathrm{L}}}\left(\frac{R_{M}}{2}(x_{n}+1)\right)\right)=0.
\label{eq:schrodinger_pot_local}
\end{eqnarray}

When a nonlocal potential is considered an additional development taking advantage of
the Chebyshev basis is pursued. As for the local potential, the nonlocal potential is also
redefined as 
\begin{eqnarray}
\mathcal{V}^{\mathrm{NL}}(r,r')=\frac{2\mu}{\hbar^{2}}V(r,r'),
\end{eqnarray}
and the following term

\begin{eqnarray}
  -\frac{R_{M}}{2}\negthickspace\int_{-1}^{+1}\negthickspace\negthickspace\negthickspace\mathcal{V^{\mathrm{NL}}}\left[\negthickspace\frac{R_{M}}{2}(x_{n}+1),\frac{R_{M}}{2}(y+1)\negthickspace\right]u\left[\negthickspace\frac{R_{M}}{2}(y+1)\negthickspace\right]dy, \nonumber \\
 \label{eq:schrodinger_terme_pot_non_local}
\end{eqnarray}
must be added to the Eq.~\eqref{eq:schrodinger_pot_local} to obtain
a Schr\"{o}dinger equation with local and nonlocal potentials. In order
to perform the integration, the potential is expressed in terms of Chebyshev
polynomials. This development concerns the variable $y$ of this potential
\begin{eqnarray}
\mathcal{V}^{\mathrm{NL}}\negthickspace\left[\negthickspace\frac{R_{M}}{2}(x+1),\frac{R_{M}}{2}(y+1)\negthickspace\right]\negthickspace\simeq\negthickspace\sum_{m=0}^{M}\overset{/}{\vphantom{a}}\negthickspace v_{m}[\frac{R_{M}}{2}(x+1)]T_{m}(y). \nonumber \\
\end{eqnarray}
The $v_{m}$ terms are obtained by considering the $M+1$ zeros of
the Chebyshev polynomial $T_{M+1}$
\begin{eqnarray}
y_{p}=\cos\frac{(p-1/2)\pi}{M+1},\quad p=1,\ldots,M+1,
\end{eqnarray}
and using the discrete orthogonality relations \cite{masson_03}, one obtains the $v_{m}$ terms,
\begin{eqnarray}
 & &\sum_{p=1}^{M+1}\mathcal{V}^{\mathrm{NL}}\left[\frac{R_{M}}{2}(x+1),\frac{R_{M}}{2}(y_{p}+1)\right]T_{m}(y_{p}) \nonumber\\
 &=& \sum_{p=1}^{M+1}\sum_{i=0}^{M}\overset{/}{\vphantom{a}}v_{i}\left[\frac{R_{M}}{2}(x+1)\right]T_{i}(y_{p})T_{m}(y_{p}) \nonumber\\
 &=& \frac{M+1}{2}v_{m}\left[\frac{R_{M}}{2}(x+1)\right].
\end{eqnarray}
The calculation of the integral in Eq.~\eqref{eq:schrodinger_terme_pot_non_local}
is thus reduced to a sum
\begin{eqnarray}
 & & \int_{-1}^{+1}\sum_{m=0}^{M}\overset{/}{\vphantom{a}}v_{m}\left[\frac{R_{M}}{2}(x+1)\right]T_{m}(y)\sum_{k=0}^{N-1}\overset{/}{\vphantom{a}}C_{k}T_{k}(y)dy \nonumber\\
 &=& \sum_{m=0}^{M}\overset{/}{\vphantom{a}}v_{m}\left[\frac{R_{M}}{2}(x+1)\right]\sum_{k=0}^{N-1}\overset{/}{\vphantom{a}}C_{k}\int_{-1}^{+1}T_{m}(y)T_{k}(y)dy \nonumber\\
 &=&\! -\!2\!\!\sum_{k=0}^{N-1}\overset{/}{\vphantom{a}}C_{k}\!\!\sum_{m=0}^{M}\overset{/}{\vphantom{a}}v_{m}\!\left[\!\frac{R_{M}}{2}(x+1)\right]\nonumber\\
 &\times&\left.\!\!\frac{m^{2}+k^{2}-1}{(m^{2}+k^{2}-1)^{2}-4m^{2}k^{2}}\right\vert _{m+k\ \mathrm{even}}.
\end{eqnarray}

The $N-2$ equations needed to determine the coefficients $C_{k}$
with local and nonlocal potentials are therefore written with $n=1,\ldots,N-2$
\begin{eqnarray}
&&\frac{4}{R_{M}^{2}}\sum_{k=2}^{N-1}C_{k}\sum_{\substack{m=0\\k-m\ \mathrm{even}}}^{k-2}\overset{/}{\vphantom{a}}(k-m)k(k+m)T_{m}(x_{n})\nonumber \\
&+&R_{M}\sum_{k=0}^{N-1}\overset{/}{\vphantom{a}}C_{k}\sum_{m=0}^{M}\overset{/}{\vphantom{a}}v_{m}\left(\frac{R_{M}}{2}(x_{n}+1)\right) \nonumber\\
&\times&\left.\frac{m^{2}+k^{2}-1}{(m^{2}+k^{2}-1)^{2}-4m^{2}k^{2}}\right\vert _{m+k\ \mathrm{even}}\nonumber\\
&+&\sum_{k=0}^{N-1}\overset{/}{\vphantom{a}}C_{k}T_{k}(x_{n})\left(\frac{2\mu E}{\hbar^{2}}-\mathcal{V^{\mathrm{L}}}\left(\frac{R_{M}}{2}(x_{n}+1)\right)\right)=0.\nonumber\\\label{eq:schrodinger_pot_local_non_local}
\end{eqnarray}

The $C_{k}$ coefficients with $k=0,..,N-1$ are calculated by solving
a system of $N$ linear equations using the LU decomposition algorithm \cite{numrec}.
The function has been modified to allow calculation with complex numbers.
Since the energy-dependent part of the phenomenological potential is
separated from the spatial part, the decomposition of the potential on
the Chebyshev polynomials needs to be performed only once for each nucleus.
Up to 100 MeV, 30 Chebyshev polynomials are sufficient to accurately describe
the wave function. Up to 250 MeV, the function must be developed over 50 polynomials.

\bibliography{references}

\end{document}